\pdfoutput=1 
\documentclass{JINST}
\usepackage{graphicx}
\usepackage{amsmath,amsthm}
\usepackage{url}
\usepackage[mediumspace,mediumqspace,squaren]{SIunits}
\usepackage{subfigure}
\usepackage{makerobust}
\MakeRobustCommand\subref

\title{\bf Effect of water vapor on the performance of glass RPCs in avalanche mode operation}

\author{K. Raveendrababu$^{a,d}$, P. K. Behera$^{a}$, B. Satyanarayana$^{b}$, S. Mukhopadhayay$^{c}$ and N. Majumdar$^{c}$ \\
\llap{$^{a}$}Physics Department, Indian Institute of Technology Madras, Chennai 600036, Tamil Nadu, India \\
\llap{$^{b}$}Department of High Energy Physics, Tata Institute of Fundamental Research, Mumbai 400005, India \\
\llap{$^{c}$}Applied Nuclear Physics Division, Saha Institute of Nuclear Physics, 1/AF, Bidhan Nagar, Kolkata 700064, India \\
\llap{$^{d}$}Physical Science Division, Homi Bhabha National Institute, Anushaktinagar, Mumbai 400094, India \\
E-mail: \email{ravi2ramana@gmail.com}}

\abstract{
We studied the effect of water vapor on the performance of glass Resistive Plate Chambers (RPCs) in the avalanche mode operation. Controlled 
and calibrated amount of water vapor was added to the RPC gas mixture that has C$_2$H$_2$F$_4$ as the major component. The deterioration in 
the performance of RPC was observed while operating with wet gas and recovered after switching to standard gas.
}

\keywords{Particle tracking detectors (Gaseous detectors), Resistive plate chambers; Large detector systems for particle and astroparticle 
physics}

\begin{document}

\maketitle                 

\section{Introduction}
\label{sec:Introduction}

The India-based Neutrino Observatory (INO) collaboration has proposed to build a 50 kiloton magnetized Iron Calorimeter (ICAL) to precisely 
measure the atmospheric neutrino oscillation parameters. The collaboration has chosen 2 $\times$ 2 m$^2$ size glass Resistive Plate Chambers 
(RPCs) as the active detector elements and is going to deploy 28,800 of them in the ICAL detector \cite{1}. The RPCs will be operated in the 
avalanche mode with an optimized gas mixture of C$_2$H$_2$F$_4$/iso-C$_{4}$H$_{10}$/SF$_{6}$ = 95.2/4.5/0.3 \cite{2}. The experiment is 
expected to run for more than 10 years in order to record statistically significant number of neutrino interactions for the confirmation of 
atmospheric neutrino oscillation. Therefore, long term stability and performance of the RPCs over the duration of the experiment is of prime 
concern. 

About 200,000 liters of gas is going to be circulating in the RPCs during the experiment. The gas lines running into about 135km in total 
are going to supply/receive gas to/from the RPC detectors. In spite of stringent QC during the RPC gas gap making or gas lines plumbing, 
it is impossible to prevent ambient air or water vapor entering into the gas circuit over these long periods of time. The contaminants are 
known cause for serious degradation in the performance or permanent damage of the RPCs \cite{3,4,5}.

Considering the severe repercussions in the mammoth ICAL detector, a systematic study of this problem was undertaken. Two glass RPCs of 
30 $\times$ 30 cm$^2$ size were fabricated and were simultaneously operated with standard gas mixture in one and with gas mixture along 
with controlled and calibrated amount of water vapor in the other. A common cosmic ray muon telescope was set up for studying both the RPCs. 
Ambient parameters such as temperature and relative humidity (RH) as well as RPCs operating and performance parameters such as currents, 
efficiencies, singles rates, signal charges and time resolutions for cosmic ray muon detection were systematically recorded throughout the 
experiment. It was observed significant deterioration in the performance of RPC in which gas with water vapor was flown.

\section{Experimental setup}
\label{sec:Experimental setup}

The RPCs of 30 $\times$ 30 cm$^2$ size were fabricated using 3 mm thick float glass plates. The outer surfaces of the glass plates were 
coated with specially developed conductive graphite paint for the ICAL RPCs, which facilitates applying high voltage across the electrodes 
\cite{6,7}. Readout strips of 2.8 cm width were orthogonally mounted on the external surfaces of the RPCs by keeping 0.2 cm gap between the 
consecutive strips. The electrode and the readout strip are separated using a layer of mylar insulator.

\begin{figure}[ht]
        \centering
         \includegraphics[width=1\textwidth]{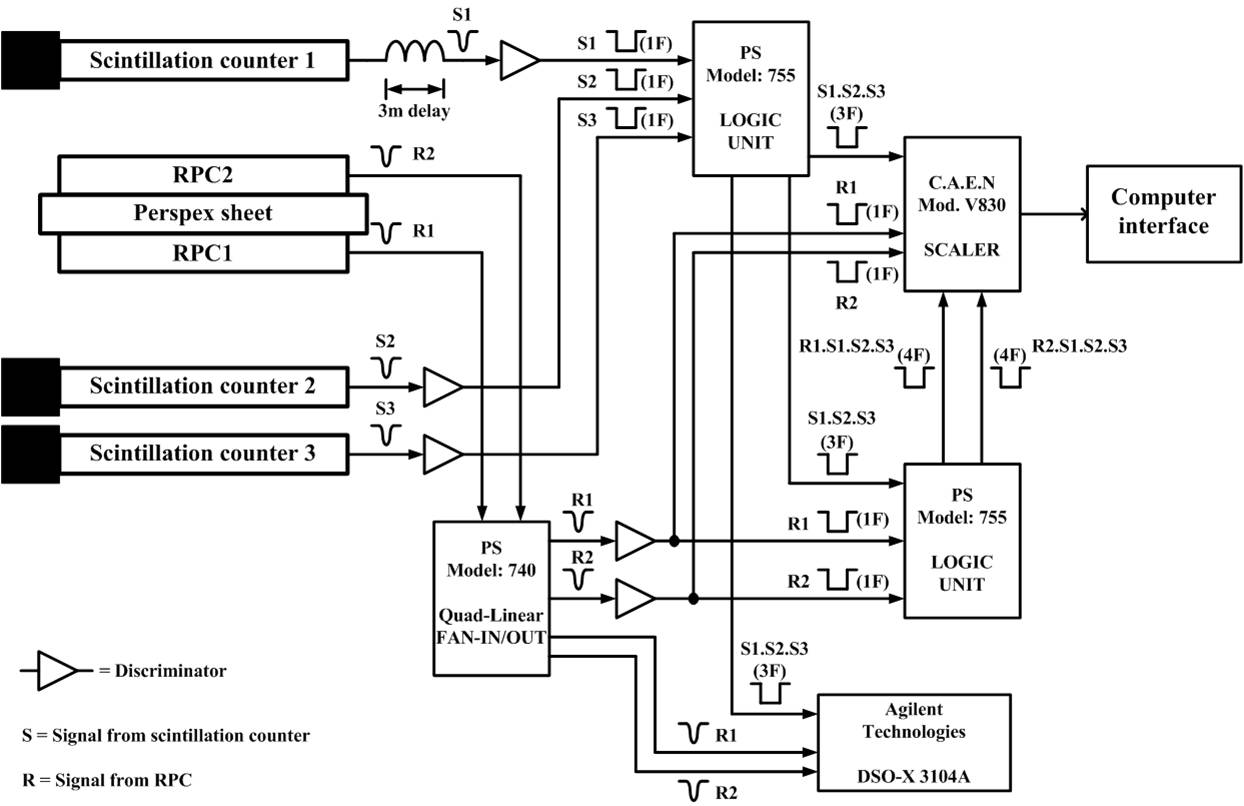}
          \caption {Block diagram of the experimental setup developed for the studies.}
\label{Fig:1}
\end{figure}

\begin{figure}[ht]
        \centering
         \includegraphics[width=0.85\textwidth]{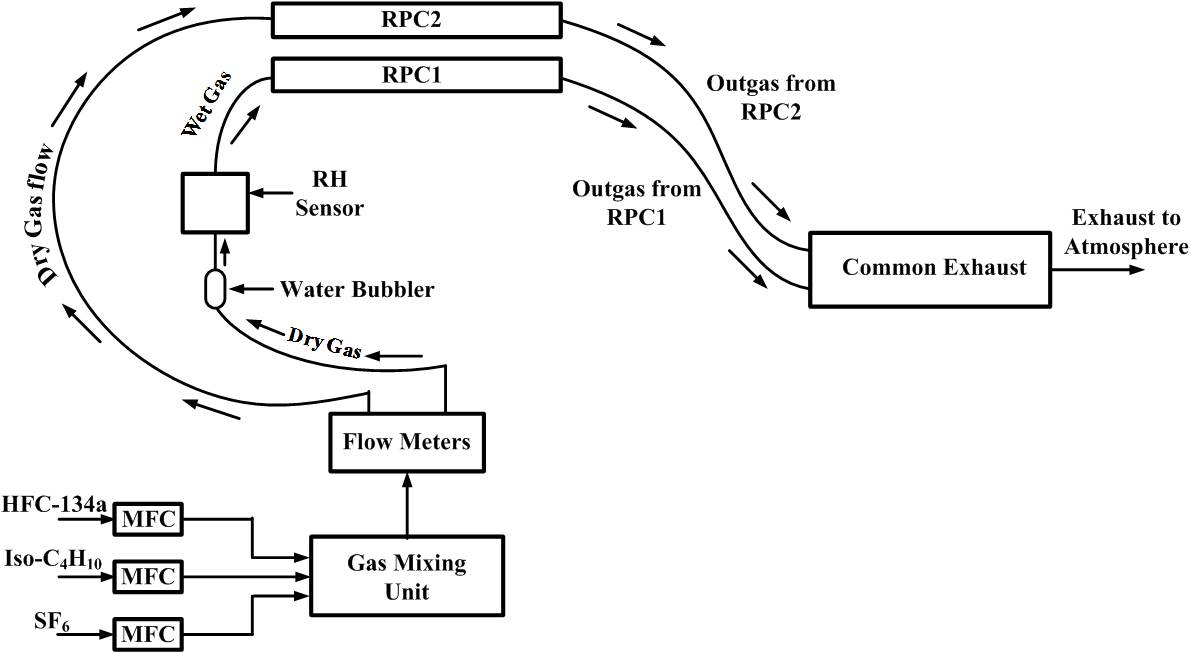}
          \caption {Block diagram of the gas flow system.}
\label{Fig:2}
\end{figure}

A telescope was set up with three plastic scintillation counters to get a 3-fold (3F) coincidence for atmospheric muons. The dimensions 
of scintillation counters in length $\times$ width $\times$ thickness are 30 cm $\times$ 2 cm $\times$ 1 cm (top), 
30 cm $\times$ 3 cm $\times$ 1 cm (middle), and 30 cm $\times$ 5 cm $\times$ 1 cm (bottom). The block diagram of a stack of two RPCs and three 
telescope counters, and the electronic circuit setup is shown in Figure~\ref{Fig:1}. An optimized gas mixture of C$_2$H$_2$F$_4$/
iso-C$_{4}$H$_{10}$/SF$_{6}$ = 95/4.5/0.5 was flown through the RPCs using polyurethane tubes. The block diagram of the gas flow system is 
shown in Figure~\ref{Fig:2}.

\section{Measurements and observations}
\label{sec:Results}  

\subsection{Standard gas studies}
\label{sec:Standard gas studies}
{
The RPCs were operated with the standard gas at 10 SCCM rate and their currents, efficiencies and singles rates as a function of applied 
voltage were measured. The measured efficiencies of RPC1 and RPC2 as a function of applied voltage are shown in Figure~\ref{Fig:3}. The 
detectors showed more than 95\% efficiency on the plateau. The measured signal charges and time resolutions of the detectors are shown in 
Figure~\ref{Fig:4}.  The signal charges of RPC1 and RPC2 are 0.99 pC and 1.24 pC, and the time resolutions are 2.8 ns and 2.3 ns, respectively. 

\begin{figure}[ht]
        \centering
         \includegraphics[width=0.5\textwidth]{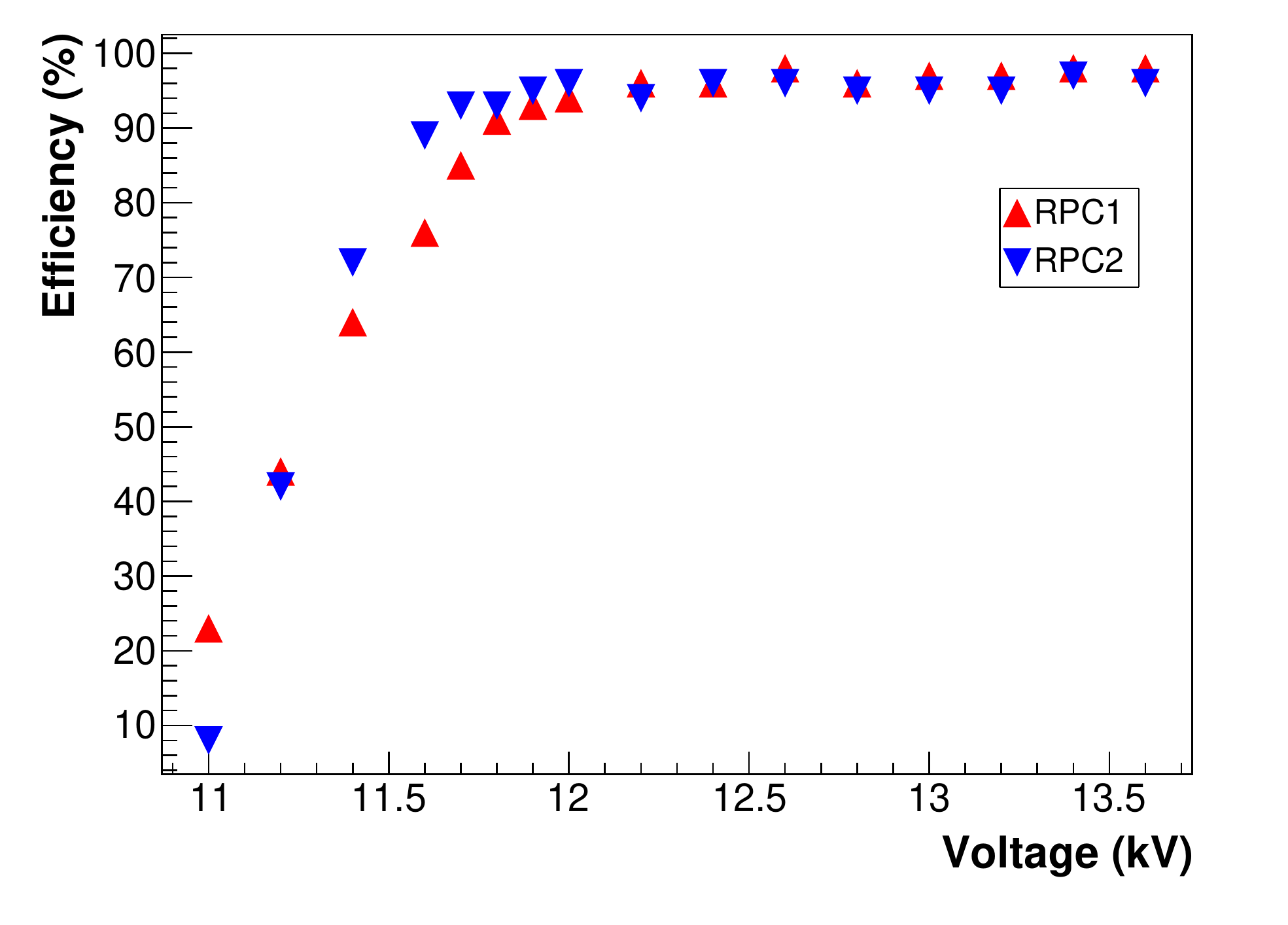}
          \caption {Efficiencies of RPC1 and RPC2 as a function of applied voltage.}
\label{Fig:3}
\end{figure}

\begin{figure}[ht]
        \centering
         \includegraphics[width=0.8\textwidth]{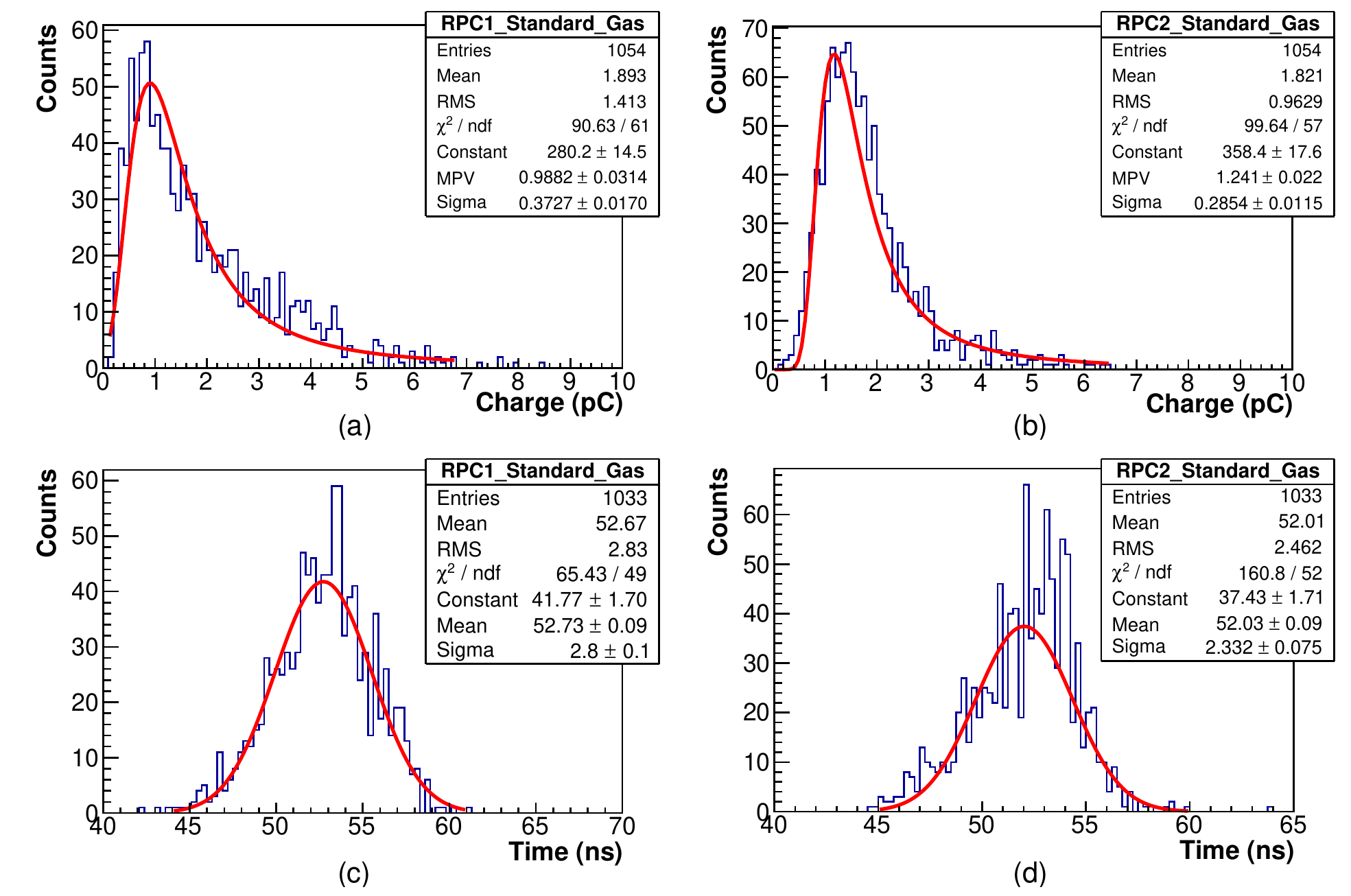}
          \caption {Signal charges: (a) and (b), and time resolutions: (c) and (d) of RPC1 and RPC2, respectively, with standard gas operation
                    at 12.2 kV.}
\label{Fig:4}
\end{figure}

\noindent {\bf Long-term stability studies:} \\
Then, the RPCs were operated at 12.2 kV and monitored their performances for 32 days. The efficiencies, singles rates and leakage currents 
of the detectors for this time period are shown in Figure~\ref{Fig:5}. Their performances were stable throughout the period.  

\begin{figure}[ht]
        \centering
         \includegraphics[width=1\textwidth]{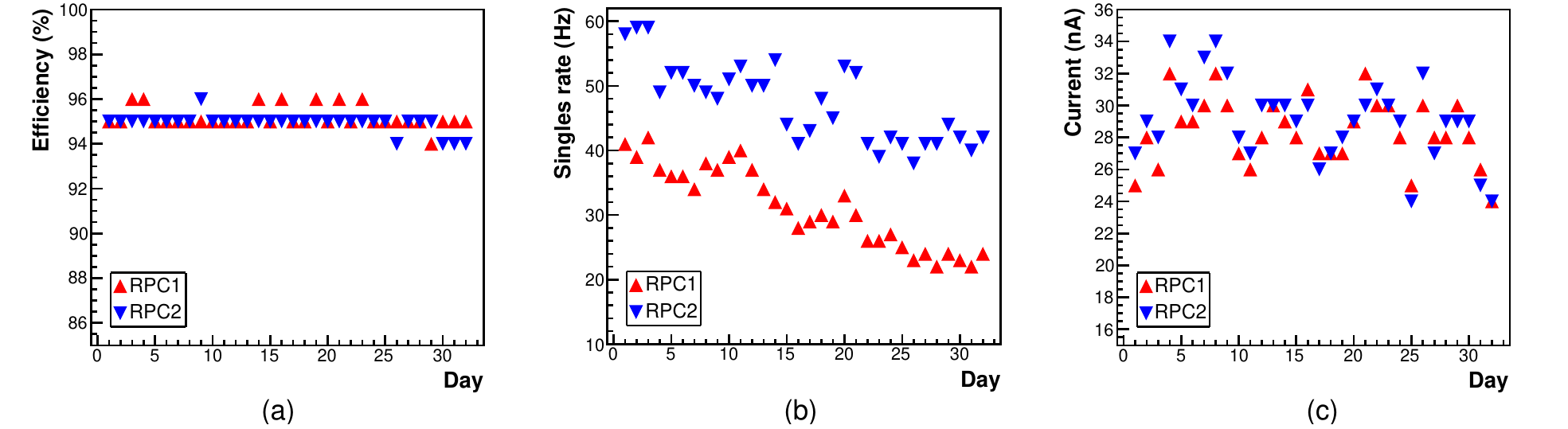}
          \caption {(a) Efficiencies, (b) singles rates and (c) currents of both the RPCs with standard gas operation for 32 days.}
\label{Fig:5}
\end{figure}
}

\subsection{Wet gas studies}
\label{sec:Wet gas studies}
{
The water vapor was started adding into RPC1 via the gas mixture using a water bubbler as shown in Figure~\ref{Fig:2}. The amount of water
vapor added to the detector was measured in a controlled way using RH sensor and is shown in Figure~\ref{Fig:6}. Throughout these studies, 
RPC2 was continued with the standard gas flow itself and was a reference detector. Then, the performances of both the detectors were monitored 
continuously.    

\begin{figure}[ht]
        \centering
         \includegraphics[width=0.5\textwidth]{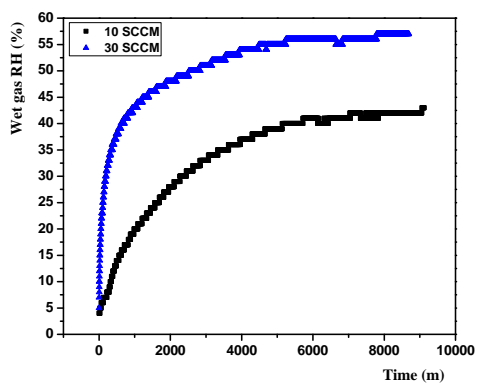}
          \caption {The quantity of water vapor addition to the RPC1 for 10 SCCM and 30 SCCM gas flow rates.}
\label{Fig:6}
\end{figure}

With the wet gas operation for a few days, the efficiency and singles rate of RPC1 degraded to 0\% and 1 Hz, respectively. The currents drawn 
by the detector increased gradually with time. The efficiency, singles rate and current drawn by the RPC1 during the wet gas operation in 
comparison to RPC2 operated with the standard gas is shown in Figure~\ref{Fig:7}. 

\begin{figure}[ht]
        \centering
         \includegraphics[width=1\textwidth]{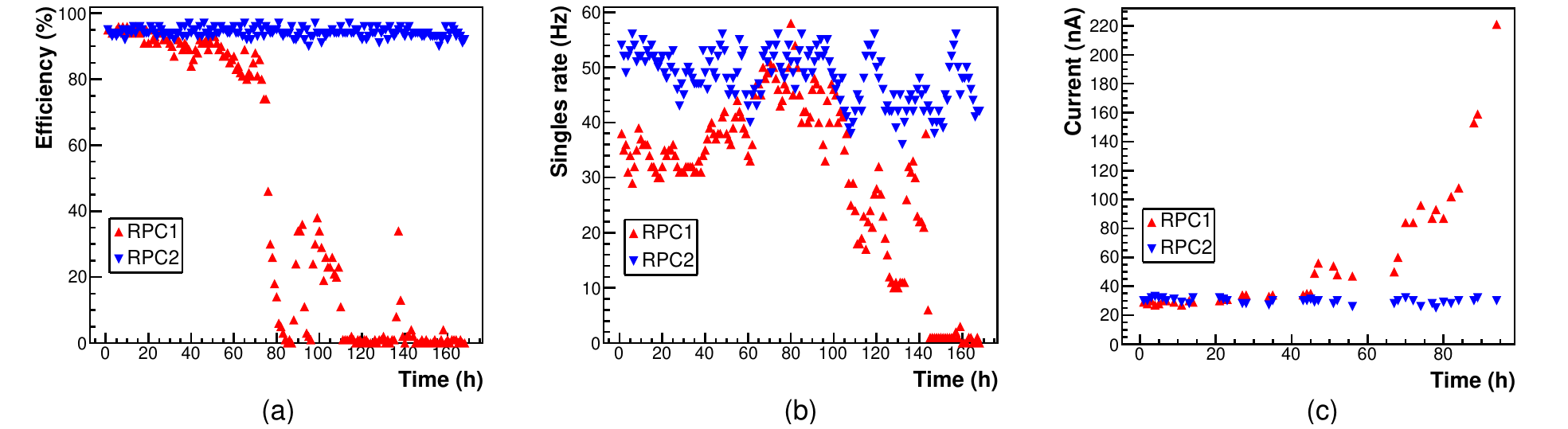}
          \caption {The (a) efficiencies, (b) singles rates and (c) currents of RPC1 with the wet gas operation, and that of RPC2 with 
                    the standard gas operation.}
\label{Fig:7}
\end{figure}

\vskip 0.5cm

\noindent
{\bf Signal charge:} \\
The signal charge of RPC1 became smaller (0.36 pC) with the wet gas operation, whereas that of RPC2 with standard gas operation (1.19 pC) 
remained similar to its measured value shown in Figure~\ref{Fig:4}b (1.24 pC). The signal charge distributions of RPC1 with wet 
gas and RPC2 with standard gas operations are shown in Figures~\ref{Fig:8}a and~\ref{Fig:8}b, respectively.  

\vskip 0.5cm

\noindent
{\bf Time resolution:} \\
The timing distribution of RPC1 with the wet gas operation got deteriorated as shown in Figure~\ref{Fig:8}c. Whereas, that of RPC2 
with standard gas operation (1.9 ns) remained similar to its measured value shown in Figure~\ref{Fig:4}d (2.3 ns).

\begin{figure}[ht]
        \centering
         \includegraphics[width=0.8\textwidth]{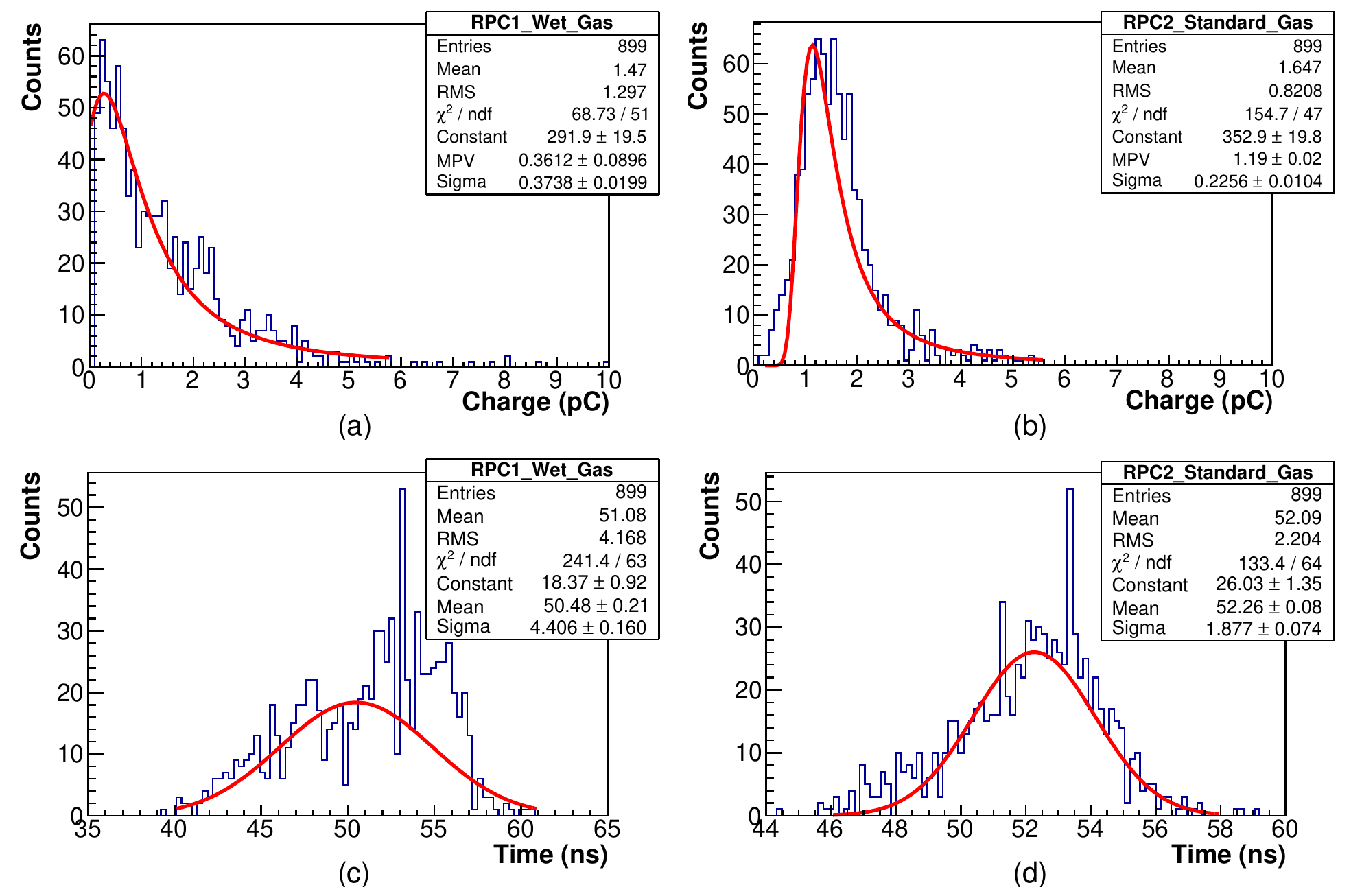}
          \caption {Signal charge of: (a) RPC1 with the wet gas operation and (b) RPC2 with the standard gas operation, and time resolution of: 
                                      (c) RPC1 with the wet gas operation and (d) RPC2 with the standard gas operation.}
\label{Fig:8}
\end{figure}
}

\subsection{Recovery studies}
\label{sec:Recovery studies}
{
RPC1 was operated with wet gas at various flow rates and the deterioration in the detector's efficiency was observed as a function of time. 
It was observed that higher the flow rate, faster the deterioration as shown in Figure~\ref{Fig:9}a. The detector was operated with wet gas 
at 0\% efficiency for a day and then switched to standard gas. Then, the efficiency of detector recovered to more than 95\% as shown in 
Figure~\ref{Fig:9}b.

\begin{figure}[ht]
        \centering
         \includegraphics[width=1\textwidth]{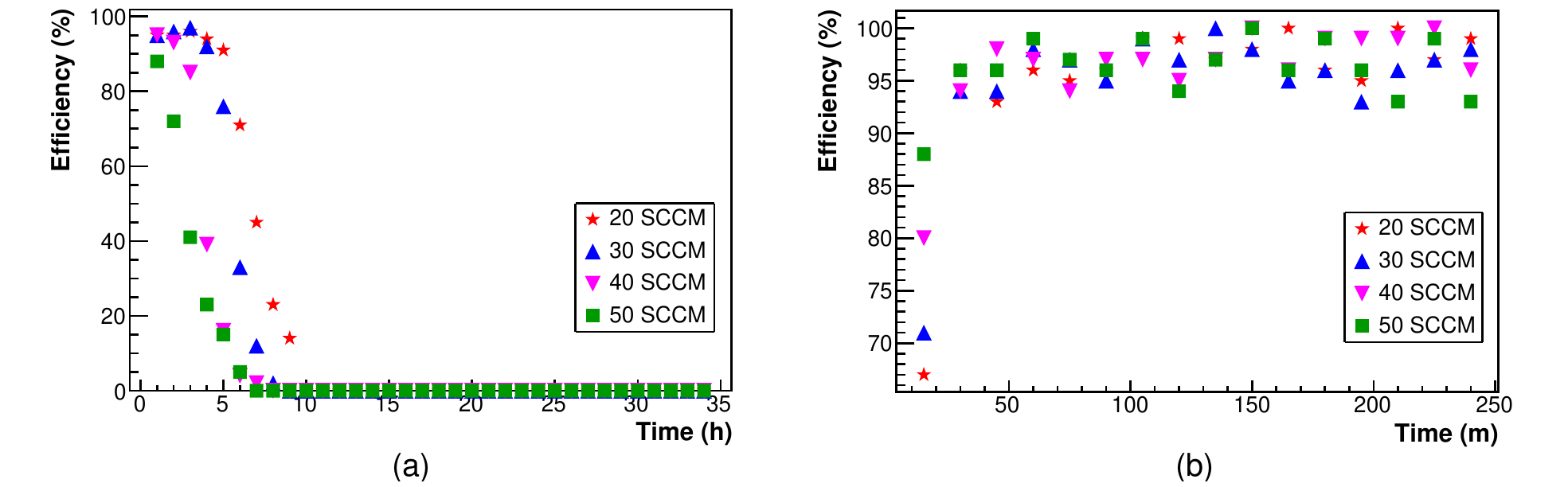}
          \caption {(a) The deterioration in efficiency of RPC1 with wet gas operation at various flow rates and (b) its recovery with
                        standard gas operation.}
\label{Fig:9}
\end{figure}
}

\section{Conclusions}
\label{sec:Conclusions}

The effect of water vapor on the performance of glass RPC was studied. We used optimized gas mixture C$_2$H$_2$F$_4$/iso-C$_{4}$H$_{10}$/
SF$_{6}$ = 95/4.5/0.5 to operate the RPC in the avalanche mode. The detector showed more than 95\% efficiency with the standard gas operation, 
its signal charge and time resolutions were 0.99 pC and 2.8 ns, respectively. Then, we started adding water vapor into the RPC. The detector's 
efficiency deteriorated to 0\% after a few days of operation with wet gas. Its signal charge became 0.36 pC and timing distribution got 
deteriorated. The RPC was operated at 0\% efficiency with wet gas for a day and then switched to standard gas. The detector's efficiency 
recovered to more than 95\%.

\section{Acknowledgements}
\label{sec:Acknowledgements}

This work was supported by the Department of Atomic Energy (DAE), and the Department of Science and Technology (DST), Government of India. 
The authors would like to gratefully acknowledge the help of their colleagues V. Asgolkar, S. Chavan, R. R. Shinde, P. Verma, S. D. Kalmani 
and L. V. Reddy at TIFR, Mumbai, and V. Janarthanam and Jafar Sadiq at IIT Madras. K. Raveendrababu expresses gratitude to P. Fonte of LIP,
Coimbra for useful discussions during the RPC2016 workshop.


\begin{thebibliography}{25}
\bibitem{1} Shakeel Ahmed et al., Physics Potential of the ICAL detector at the India-based Neutrino Observatory (INO), arXiv:1505.07380v1.
\bibitem{2} M. Salim et al., Experimental and numerical studies on the effect of SF$_{6}$ in a glass RPC, 2012 JINST 7 P11019.
\bibitem{3} H. Sakai et al., Study of the effect of water vapor on a resistive plate chamber with glass electrodes, 
            Nucl. Instrum. and Meth. A 484 (2002) 153-161.
\bibitem{4} T. Kubo et al., Study of the effect of water vapor on a glass RPC with and without freon, 
            Nucl. Instrum. and Meth. A 508 (2003) 50-55.
\bibitem{5} A. Candela et al., Ageing and recovering of glass RPC, Nucl. Instrum. and Meth. A 533 (2004) 116-120.  
\bibitem{6} R. Santonico and R. Cardarelli, Development of Resistive Plate Counters, Nucl. Instrum. and Meth. 187 (1981) 377-380.
\bibitem{7} M. Bhuyan et al., Development of 2 m $\times$ 2 m size glass RPCs for INO, Nucl. Instrum. and Meth. A 661 (2012) S64-S67.

\end{thebibliography}
\end{document}